\documentclass[twocolumn,american]{article}
\usepackage[T1]{fontenc}
\usepackage[latin9]{inputenc}
\usepackage{babel}
\usepackage{verbatim}
\usepackage{units}
\usepackage{amsmath}
\usepackage{amssymb}
\usepackage{graphicx}
\usepackage[pdfusetitle,
 bookmarks=true,bookmarksnumbered=false,bookmarksopen=false,
 breaklinks=false,pdfborder={0 0 1},backref=false,colorlinks=false]
 {hyperref}

\makeatletter

\providecommand{\tabularnewline}{\\}

\@ifundefined{date}{}{\date{}}
\usepackage[left=2cm,right=1.5cm,top=1cm,bottom=1cm,includeheadfoot]{geometry}
\makeatother

\begin{document}
\title{A solvable microscopic model for the propagation of light in a dielectric
medium}
\author{R. Dengler \thanks{ORCID: 0000-0001-6706-8550}\\
Munich, Germany}
\maketitle
\begin{abstract}
Maxwell's equations resemble Schr\"{o}dinger's equation in that an exact
solution for a well-defined model delivers all physically relevant
details. Solvable microscopic electrodynamic models, however, are
rare. An exception is the discrete dipole approximation (DDA), which
models a medium as a lattice of point dipoles. We use a regularized
DDA variant to examine mechanical and electromagnetic momentum of
light signals in such a medium in detail. The results agree in essential
parts with that of the theory of R. Peierls from 1976.
\end{abstract}

\section{Introduction}

Exact solutions of idealized models derived from first principles
are fundamental to theoretical physics. In the context of light propagation
in microscopic dielectric media, such solutions are known; however,
their detailed implications within the medium have not, to our knowledge,
been systematically investigated.

\begin{figure}
\centering{}\includegraphics[scale=1.3]{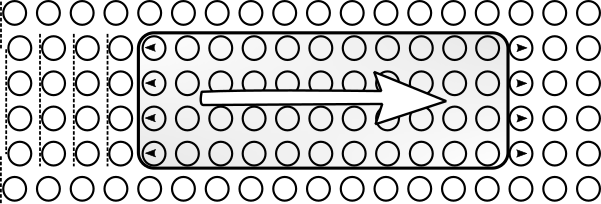}\caption{\label{fig:Signal}A light signal in a solid. The signal (the gray
block) moves to the right, slightly accelerates atoms at its right
front to the right, and slightly decelerates atoms at its left front,
thus leaving behind a trace of atoms shifted to the right, but again
at rest. Atoms within the signal are moving to the right and contribute
to the signal momentum.}
\end{figure}

The physical context is depicted in figure~\ref{fig:Signal}. A wave
packet of light enters and traverses a dielectric slab of thickness
$L$. The wave packet is much shorter than $L$, making its propagation
analogous to that of a classical particle. This \textquotedbl particle\textquotedbl{}
carries a well-defined amount of energy and momentum. It is well known
that a significant portion of the momentum is mechanical in nature.
What is going on is explained in~\cite{Gordon1973,Peierls1976} and
also shown in figure~\ref{fig:Signal}. The question of the amount
of this momentum and its relation to the Abraham and Minkowski expressions
for momentum density, has been the subject of a long-standing debate.
A physically often more directly relevant quantity is the momentum
flow, the longitudinal stress tensor, but this does not answer the
original question.

This is not the place to review the arguments or the current state
of this broad and diverse field. Instead, we refer readers to existing
reviews~\cite{Brevik1979,Kemp2011,Partanen_2022} and recent contributions~\cite{Barn2010,BarnLoud2010,Silveir2017,Partanen_2017,Partanen_2023}.
We will revisit some aspects later. Clearly the problem is complex
in many respects. As often in the case of controversies or complications,
exact solutions can offer valuable new insights.

The specific physical situation described above justifies a crucial
simplification~\cite{Peierls1976,Peierls1977}. Order of magnitude
considerations confirm that the atoms remain effectively stationary
while the wave packet traverses the slab. Consider a light signal
with an energy density $u$. Equating the electromagnetic momentum
$u\ell^{3}/c$ in a cube of volume $\ell^{3}$ with a mechanical momentum
$\mu\ell^{3}v$ yields an equation for the velocity $v$ of matter
in terms of mass density $\mu$. The resulting displacement of the
atoms in time $\ell/c$ is

\begin{equation}
\Delta x\sim u\ell/\left(\mu c^{2}\right).\label{eq:Shift_Estimate}
\end{equation}
For a light signal with a power of $\unit[15]{mW}$ in a cross section
of $\ell^{2}=\left(\unit[0.6]{mm}\right)^{2}$ in a medium with $\mu=10^{3}\unit{kg/m^{3}}$
this leads to $\Delta x\sim\unit[10^{-27}]{m}$ (the values are from
the experiment of Jones and Leslie~\cite{Jones_Leslie_1978}). For
signals that persist over longer times $t$, the atoms remain in motion
for a correspondingly longer duration, and the displacement grows
linearly with the signal length (or time), scaled by a factor $ct/\ell$
. For $\ell=\unit[0.6]{mm}$ the factor is approximately $10^{11}/\unit{s}$.
Even after an hour, the resulting displacement remains negligible.
The same holds for the kinetic energy of the atoms - and, of course,
no electrostriction occurs.

Thus, although generic modes or quasiparticles do couple electromagnetic
and mechanical degrees of freedom, in most practical situations, it
is sufficient to assume that the atoms absorb momentum while effectively
remaining stationary.

\section{The model}

A simple microscopic model of a dielectric medium consists of atoms
treated as classical oscillators (akin to the Lorentz model) arranged
on a cubic lattice with spacing $a$. The atoms have a finite radius
$R$. In the limit $R\rightarrow0$, this reduces to the discrete
dipole approximation~\cite{Draine_1988,Draine_1993,Yurkin_2007},
a formalism commonly used to study light scattering by arbitrarily
shaped dust particles. Unlike the DDA, however, to avoid singularities
we retain a finite atomic radius. And we only consider translationally
invariant systems, which allow an exact analytic solution~\cite{Draine_1993}.

\subsection{Basic idea of the discrete dipole approximation}

The starting point is the inhomogeneous Helmholtz equation for the
electric field $E$, in SI units  

\begin{align}
\left(\nabla^{2}+\omega^{2}\right)E_{m} & =\tfrac{1}{\epsilon_{0}}\left(\nabla_{m}\rho-i\omega j_{m}\right).\label{eq:Helmholtz3D}
\end{align}
For simplicity we set the speed of light in vacuum to unity ($c=1$).
This equation follows directly from the Maxwell equations assuming
a factor $e^{-i\omega t}$ for the fields, the charge density $\rho$
and the current density $j$. We usually omit the factor $e^{-i\omega t}$
in the following. The source terms on the r.h.s. of Eq.~(\ref{eq:Helmholtz3D})
are different from zero only in the polarizable matter, and in general
also depend on the fields.

A kind of decoupling occurs if the atoms have a radius $R\ll a\ll1/\omega$.
The interaction with the atoms at the lattice sites $x_{m}$ then
is due to an electric field $E^{\mathrm{loc}}e^{ikx_{m}}$ containing
a constant vector $E^{\mathrm{loc}}$ and the Bloch factor $e^{ikx_{m}}$.
The smallness of the atoms means that the electric field in their
vicinity is the quasi static near field of electric dipole radiation,
related to $\rho$ and $j$ by an electrostatic problem. In other
words, near the atoms the derivative $\nabla^{2}\sim R^{-2}$ dominates
the $\omega^{2}$ in the Helmholtz equation. The strategy then is
to first solve the electrostatic problem, express the r.h.s. of the
Helmholtz equation~(\ref{eq:Helmholtz3D}) in terms of $E^{\mathrm{loc}}$,
and then solve the resulting linear equation in Fourier space. 

\subsection{Classical atoms}

As a classical (non-quantum-mechanical) model for an atom we use two
homogeneously charged rigid spheres of radius $R$, one with charge
$-Q$, and another \emph{heavier} sphere with charge $+Q$. The spheres
can interpenetrate each other, initially they overlap. Rigid spheres
are not compatible with special relativity, however, the involved
speeds are small, and no problems arise. The force acting between
the spheres is due to the electric force attracting the spheres and
an additional potential energy, and can be made harmonic with any
desired force constant by choosing the additional potential energy
appropriately. Such an atom resembles a dielectric sphere for which,
however, the electric susceptibility would be limited by the polarization
catastrophe.

The negatively charged spheres have negligible mass and follow any
electric field in phase. Under the influence of a local electric field
$E^{\mathrm{loc}}$ (not including the field of the atom itself) an
atom acquires a dipole moment

\begin{equation}
p^{\left(0\right)}=\epsilon_{0}\gamma a^{3}E^{\mathrm{loc}}=Qb,\label{eq:Dip_Moment}
\end{equation}
where $\gamma$ is the (dimensionless) susceptibility and $b$ the
displacement. We consider the limit of large $Q$ and small $b$.
The displacement $b$ generates a charge density on the surface of
the atom. The corresponding depolarization field within the sphere
has the well-known value~\cite{Jackson1975}
\begin{equation}
E_{P}=-p^{\left(0\right)}/\left(3\epsilon_{0}V_{R}\right),\label{eq:E_Depol}
\end{equation}
where $V_{R}$ is the volume of the sphere. This field generates an
attractive force
\[
F_{P}=E_{P}Q=-Qp^{\left(0\right)}/\left(3\epsilon_{0}V_{R}\right)\equiv-\lambda_{P}b,
\]
 between the spheres with spring constant $\lambda_{P}=Q^{2}/\left(3\epsilon_{0}V_{R}\right).$

The polarization law (\ref{eq:Dip_Moment}), on the other hand, can
be written as $b=\left(\epsilon_{0}\gamma a^{3}/Q^{2}\right)QE^{\mathrm{loc}}$,
and thus implies a spring constant $\lambda_{\mathrm{tot}}=Q^{2}/\left(\epsilon_{0}\gamma a^{3}\right),$
smaller  than $\lambda_{P}$ by the factor 
\begin{equation}
\lambda_{\mathrm{tot}}/\lambda_{P}=3V_{R}/\left(\gamma a^{3}\right).\label{eq:Spring_Ratio}
\end{equation}
In other words, to get the susceptibility $\gamma$ one needs an additional
potential energy. This plays a role if one wants to calculate the
energy of the system from the electric and the magnetic fields alone.

The induced depolarization field (\ref{eq:E_Depol}) contributes to
the electric field in the sphere and thus
\begin{equation}
E\left(x=0\right)=\left(1-\tfrac{\gamma a^{3}}{3V_{R}}\right)E^{\mathrm{loc}}.\label{eq:E_0_3D}
\end{equation}
The amplitude of this field is much larger than $E^{\mathrm{loc}}$
for $\gamma=O\left(1\right)$ and $R\ll a$.

\subsection{Electrostatics}

In the limit $R\ll a\ll\lambda$ the charge and current density on
the r.h.s. of Eq.~(\ref{eq:Helmholtz3D}) can be deduced from the
quasi static limit, that is from the near field equations, where $\nabla^{2}\sim R^{-2}$
dominates $\omega^{2}\sim\lambda^{-2}$.  Charge and current density
of the atom at the origin then take the form

\begin{align}
\rho^{\left(0\right)}\left(x\right) & =-\epsilon_{0}\gamma E^{\mathrm{loc}}\tfrac{a^{3}}{V_{R}}\cdot\nabla\theta\left(R-\left|x\right|\right),\label{eq:rho_0}\\
j^{\left(0\right)}\left(x\right) & =-i\omega\epsilon_{0}\gamma E^{\mathrm{loc}}\tfrac{a^{3}}{V_{R}}\theta\left(R-\left|x\right|\right),\label{eq:j_0}
\end{align}
where $\theta$ is the step function.

\subsubsection{Charge and current density}

The total charge and current densities are the superpositions of the
quantities~(\ref{eq:rho_0}, \ref{eq:j_0}) for all lattice points
with the additional Bloch factor $e^{ikx}$. The wavevector $k$ is
assumed to point into the $x_{3}$ direction. Of interest are the
corresponding Fourier series

\begin{align}
\rho\left(x\right) & =-i\epsilon_{0}\gamma E^{\mathrm{loc}}e^{ikx}\cdot\sum_{q\in\left(2\pi\mathbb{Z}/a\right)^{3}}e^{iqx}qF\left(qR\right),\label{eq:rho_Fourier}\\
j\left(x\right) & =-i\omega\epsilon_{0}\gamma E^{\mathrm{loc}}e^{ikx}\sum_{q\in\left(2\pi\mathbb{Z}/a\right)^{3}}e^{iqx}F\left(qR\right),\label{eq:j_Fourier}\\
F\left(s\right) & =3\left(\tfrac{\sin s}{s^{3}}-\tfrac{\cos s}{s^{2}}\right)=1-\tfrac{s^{2}}{10}+O\left(s^{4}\right).\label{eq:F_s}
\end{align}
The function $F$ is defined in Eq.~(\ref{eq:F_s_Def}) in the appendix.
It should be mentioned that simply adding the Bloch factor to Eq.~(\ref{eq:rho_0},
\ref{eq:j_0}) is only valid in the long wavelength limit. The continuity
equation for $\rho$ and $j$ is not exactly satisfied, there remains
a term of order $O\left(ka\right)$. This term is negligible in the
long wavelength limit. 

\subsubsection{Average electric field}

The continuous medium Maxwell equations use the average (macroscopic)
electric field $E^{\mathrm{avg}}$, which we now want to relate to
$E^{\mathrm{loc}}$. This relation also is a direct consequence of
the exact solution~(\ref{eq:E_x_3D}) of the Helmholtz equation below.
It may nevertheless be instructive to derive the formula in the context
of electrostatics.

The field $E^{\mathrm{loc}}$, responsible for the polarization of
the atom at the origin, does not include the field from the atom itself.
It can be identified with the field generated by the dipoles at $\left|x\right|>\tilde{R}$
where $\tilde{R}$ is some radius much larger than the lattice spacing.
This field $E^{\mathrm{loc}}$ essentially is constant in the sphere
$\left|x\right|<\tilde{R}$. The dipoles at $\left|x\right|<\tilde{R}$
do not contribute to the field at the origin according to the usual
Clausius-Mossotti calculation~\cite{Jackson1975}. However, they
also contribute to the average field. The average electric field in
a sphere with polarization $P$ (\emph{without} a polarizing external
field) is $E=-P/\left(3\epsilon_{0}\right)$~\cite{Jackson1975},
see also the similar Eq~.(\ref{eq:E_Depol}).  Combining the two
contributions and using~(\ref{eq:Dip_Moment}) yields

\begin{equation}
E^{\mathrm{avg}}=\left(1-\tfrac{\gamma}{3}\right)E^{\mathrm{loc}}.\label{eq:E_avg}
\end{equation}
For the dipole moment~(\ref{eq:Dip_Moment}) this implies $p^{\left(0\right)}=\epsilon_{0}E^{\mathrm{avg}}\gamma a^{3}/\left(1-\gamma/3\right)$,
which is the Clausius-Mossotti relation.

\section{Formal solution}

The Bloch condition allows to write the electric field as a Fourier
series,

\begin{equation}
E\left(x\right)=e^{ikx}a^{-3}\sum_{q\in\left(2\pi\mathbb{Z}/a\right)^{3}}E_{q}e^{iqx}.\label{eq:E_Fourier}
\end{equation}
The Helmholtz equation~(\ref{eq:Helmholtz3D}) can then be solved
by inserting the Fourier series~(\ref{eq:rho_Fourier}), (\ref{eq:j_Fourier})
and (\ref{eq:E_Fourier}). The result is

\begin{align}
E\left(x\right) & =\gamma\left|E^{\mathrm{loc}}\right|e^{ikx}\sum_{q\in\left(2\pi\mathbb{Z}/a\right)^{3}}e^{iqx}\tfrac{\omega^{2}\mathbf{1}-\left(q+k\right)q}{\left(q+k\right)^{2}-\omega^{2}}\hat{e}F\left(qR\right),\label{eq:E_x_3D0}
\end{align}
where $\hat{e}$ is the polarization direction and $\left(q+k\right)q$
a matrix. Eq.(\ref{eq:E_x_3D0}) agrees with the DDA solution~\cite{Draine_1993},
only the shape function $F\left(qR\right)$ is new. 

\subsection{Dispersion relation}

The dispersion relation can be deduced from a consistency condition:
the electric field~(\ref{eq:E_x_3D0}) at the origin must agree with
the field~(\ref{eq:E_0_3D}) at the center of the atom. In the long
wavelength limit Eq.~(\ref{eq:E_x_3D0}) leads to a simple expression
for $E\left(0\right)$ by first writing down the $q=0$ term of the
sum and then using the fact that $k$ and $\omega$ are negligible
against any $q\neq$0 in the remaining sum,

\begin{align}
\frac{\left|E\left(0\right)\right|}{\left|E^{\mathrm{loc}}\right|} & =\gamma\tfrac{\omega^{2}}{k^{2}-\omega^{2}}-\tfrac{\gamma}{3}\left(\sum_{q}F\left(qR\right)-1\right)\label{eq:DispRel}\\
 & \cong\gamma\tfrac{\omega^{2}}{k^{2}-\omega^{2}}-\tfrac{\gamma a^{3}}{3V_{R}}+\tfrac{\gamma}{3}.\nonumber 
\end{align}
The sum $\sum_{q}F\left(qR\right)=a^{3}/V_{R}$ directly follows from
the definition~(\ref{eq:F_s_Def}). The comparison with Eq.~(\ref{eq:E_0_3D})
leads to the dispersion relation

\begin{align}
\omega^{2} & =k^{2}/n^{2},\label{eq:Disp_3D}\\
n^{2} & =\frac{1+2\gamma/3}{1-\gamma/3}.\label{eq:n2_3D}
\end{align}
The last equation is the Clausius-Mossotti formula for the refractive
index $n$.

\subsection{Electric and magnetic field}

In the $q\neq0$ parts of the electric field~(\ref{eq:E_x_3D0})
$q\sim2\pi/a$ dominates $k$ and $\omega$ in the long wave length
limit, and one can write

\begin{equation}
E\left(x\right)=\left|E^{\mathrm{loc}}\right|e^{ikx}\left[\left(1-\tfrac{\gamma}{3}\right)\mathbf{1}-\gamma\sum_{q\neq0}e^{iqx}\tfrac{qq}{q^{2}}F\left(qR\right)\right]\hat{e}.\label{eq:E_x_3D}
\end{equation}
The $q=0$ part agrees with the average field $E^{\mathrm{avg}}$
from Eq.~(\ref{eq:E_avg}) modulo the Bloch phase.

From Eq.~(\ref{eq:E_x_3D0}) the magnetic field follows as

\begin{align}
B\left(x\right) & =\frac{\nabla\times E}{i\omega}\cong\left|E^{\mathrm{avg}}\right|e^{ikx}n\hat{k}\times\hat{e}.\label{eq:B_x_3D}
\end{align}
Here we have used the dispersion relation~(\ref{eq:Disp_3D}, \ref{eq:n2_3D}).
The contributions to $B\left(x\right)$ from $q\neq0$ contain a factor
$\omega a\rightarrow0$. In the long wavelength limit the magnetic
field thus is constant apart from the factor $e^{ikx}$. This important
fact has a generic explanation: the fields in the near zone of electric
dipole radiation (caused by the atoms) are dominantly electric in
nature~\cite{Jackson1975}. A substantial magnetic field only arises
in the far zone, the total magnetic field thus has negligible granular
structure.  

The solution~(\ref{eq:E_x_3D}, \ref{eq:B_x_3D}) depends on the
relation (\ref{eq:Dip_Moment}) between the electric field and the
dipole moment. One might object that in a plane wave there also acts
a Lorentz force on the negative charges. However, an order-of-magnitude
estimate using~(\ref{eq:B_x_3D}) shows $F^{\mathrm{lorentz}}/F^{\mathrm{coulomb}}\sim kb<ka.$
The Lorentz force is thus negligible in Eq.~(\ref{eq:Dip_Moment})
in the long-wavelength limit.

\section{Densities of energy and momentum}

Energy and momentum densities can be obtained by averaging over a
unit cell of the lattice and performing the time average. For clarity,
table~\ref{tab:VacuumQuantities} lists the corresponding expressions
for electromagnetic quantities in vacuum. 
\begin{table}
\centering{}%
\begin{tabular}{l|l}
\hline 
$\;u^{\mathrm{em}}$ & $\tfrac{\epsilon_{0}}{2}\left(E^{2}+B^{2}\right)$\tabularnewline
$\;\Pi^{\mathrm{em}}$ & $\epsilon_{0}\left(\boldsymbol{E}\times\boldsymbol{B}\right)$\tabularnewline
$\;\sigma_{i,j}^{\mathrm{em}}$ & $\tfrac{\epsilon_{0}}{2}\left(\delta_{ij}\left(E^{2}+B^{2}\right)-2E_{i}E_{j}-2B_{i}B_{j}\right)$\tabularnewline
\hline 
\end{tabular}\caption{\label{tab:VacuumQuantities}Electromagnetic energy density $u$,
momentum density $\Pi$ and stress tensor $\sigma$ in vacuum in SI
units for $c=1$. }
\end{table}

\subsection{Energy density}

The average energy density due to the electric field~(\ref{eq:E_x_3D})
is

\begin{align}
\overline{u^{\mathrm{elec}}} & =\tfrac{\epsilon_{0}}{4}a^{-3}\int_{a^{3}}\mathrm{d}^{3}x\left|E\right|^{2}\label{eq:u^elec}\\
 & =\tfrac{\epsilon_{0}}{4}\left|E_{2}^{\mathrm{loc}}\right|^{2}\left[\left(1-\tfrac{\gamma}{3}\right)^{2}-\tfrac{\gamma^{2}}{3}+\tfrac{\gamma^{2}}{3}\tfrac{a^{3}}{V_{R}}\right].\nonumber 
\end{align}
The evaluation of the integral is described in appendix~\ref{subsec:u_em_calc}.
The contribution proportional to $1/V_{R}$ is due to the electromagnetic
field of the individual dipoles, which agrees with the potential energy
of an oscillator with spring constant $\lambda_{P}$.  The actual
spring constant is smaller by a factor~(\ref{eq:Spring_Ratio}),
and the total average ``electric'' energy density becomes

\begin{align}
\overline{u^{\mathrm{elec,tot}}}= & \tfrac{\epsilon_{0}}{4}\left|E^{\mathrm{loc}}\right|^{2}\left(1-\tfrac{\gamma}{3}\right)^{2}n^{2}=\tfrac{\epsilon_{0}}{4}\left|E^{\mathrm{avg}}\right|^{2}n^{2}.\label{u_elec_tot}
\end{align}
Here we have used Eq.~(\ref{eq:n2_3D}) and the expression~(\ref{eq:E_avg})
for the average electric field $E^{\mathrm{avg}}$. The expression
$\overline{u^{\mathrm{elec,tot}}}$ in terms of $E^{\mathrm{avg}}$
agrees with the electric energy density of the Maxwell equations for
a continuous medium with refractive index $n$. This could have been
expected. The medium with waves with a long wavelength locally looks
like the medium in a capacitor, and energy and propagation speed only
depend on the capacitance, not on other details.

The magnetic contribution to the energy density directly follows from
Eq.~(\ref{eq:B_x_3D}) and has the same value as~(\ref{u_elec_tot}),
and thus as expected
\begin{equation}
\overline{u^{\mathrm{tot}}}=\tfrac{\epsilon_{0}}{2}\left|E^{\mathrm{avg}}\right|^{2}n^{2}.\label{eq:u_tot}
\end{equation}

\subsection{Electromagnetic momentum density}

We now calculate the average of the electromagnetic momentum density
$\Pi^{\mathrm{em}}=\epsilon_{0}\Re E\times\Re B$ from the fields~(\ref{eq:E_x_3D},
\ref{eq:B_x_3D}). Since the magnetic field is $q$-independent there
only contributes $E^{q=0}\left(x\right)=E^{\mathrm{avg}}e^{ikx}$
and thus
\begin{equation}
\overline{\Pi^{\mathrm{em}}}=\tfrac{\epsilon_{0}}{2}\left|E_{2}^{\mathrm{avg}}\right|^{2}n\hat{k}.\label{eq:PI_em}
\end{equation}
This means that the result~(\ref{eq:PI_em}) is generic. The electromagnetic
momentum per energy follows as $\left|\overline{\Pi^{\mathrm{em}}}\right|/\overline{u^{\mathrm{tot}}}=1/n$. 

\subsection{Momentum density from stress tensor}

\label{subsec:sigma33}To simplify the expressions we now consider
a signal propagating in $x_{3}$ direction with polarization in $x_{2}$
direction.

We want to calculate the momentum flow in propagation direction of
the plane wave with the fields~(\ref{eq:E_x_3D}, \ref{eq:B_x_3D}),
given by the integral of the stress tensor component
\begin{equation}
\sigma_{33}=\tfrac{\epsilon_{0}}{2}\left(E^{2}+B^{2}-2E_{3}E_{3}\right)\label{eq:sigma_33_def}
\end{equation}
over a surface perpendicular to the propagation direction. The average
momentum flow is the same for any surface at $x_{3}\in\left[R/2,a-R/2\right]$.
The $q=0$ components of $E$ and $B$ directly give constant contributions
in Eq.~(\ref{eq:sigma_33_def}), there remains the contribution from
$E_{q\neq0}\left(x\right)$ of the solution~(\ref{eq:E_x_3D}).

For simplicity we use $x_{3}=a/2$. Since $x_{3}=a/2$ is far away
from the atoms, one can replace the atoms with point matter, that
is $F\left(qR\right)=1$. 

Without the trivial factors $E_{2}^{\mathrm{loc}}e^{ikx}\gamma$ from
Eq.~(\ref{eq:E_x_3D}) the granular part of the electric field is
\begin{align}
\tilde{E}_{j}\left(x\right) & =\nabla_{j}\nabla_{2}\sum_{q\neq0}e^{iqx}\tfrac{1}{q^{2}}\equiv\nabla_{j}\nabla_{2}S.\label{eq:E_tilde}
\end{align}
One could evaluate this sum with the Ewald method~\cite{Born_1954},
but there is a more direct way. The sum over $q_{3}$ can be performed
in closed form. Noting that contributions to $\tilde{E}$ only arise
from the domain $Q^{2}=q_{1}^{2}+q_{2}^{2}\neq0$ one can write (see
appendix~\ref{subsec:Sum_cosh_formula})

\begin{align}
S & =\sum_{q\in\left(2\pi\mathbb{Z}/a\right)^{3}\setminus0}e^{iqx}\tfrac{1}{q^{2}}\label{eq:sum_S}\\
 & =\tfrac{a^{2}}{4}\sum_{\left\{ q_{1},q_{2}\right\} \in\left(2\pi\mathbb{Z}/a\right)^{2}\setminus0}e^{i\left(q_{1}x_{1}+q_{2}x_{2}\right)}\tfrac{\cosh\left(Q\left(x_{3}-\tfrac{a}{2}\right)\right)}{\tfrac{Qa}{2}\sinh\left(\tfrac{Qa}{2}\right)}.\nonumber 
\end{align}
For the plane at $x_{3}=a/2$ it follows

\[
\tilde{E}_{j}\left(\tfrac{a}{2}\right)=-\left(\delta_{j,1}+\delta_{j,2}\right)\tfrac{a^{2}}{4}\sum_{\left\{ q_{1},q_{2}\right\} \neq0}\tfrac{e^{i\left(q_{1}x_{1}+q_{2}x_{2}\right)}q_{j}q_{2}}{\tfrac{Qa}{2}\sinh\left(\tfrac{Qa}{2}\right)}.
\]
For the average of the stress tensor~(\ref{eq:sigma_33_def}) on
the plane we need the constant

\begin{align}
M & =a^{-2}\int_{a^{2}}\mathrm{d}x_{1}\mathrm{d}x_{2}\left|\tilde{E}\right|^{2}\label{eq:M_Madelung}\\
 & =\tfrac{1}{2}\sum_{\left\{ q_{1},q_{2}\right\} \neq0}\left(\tfrac{Qa}{2}/\sinh\tfrac{Qa}{2}\right)^{2}\cong0.1715990.\nonumber 
\end{align}
The integral over the area enforces the same wavevectors in the two
$\tilde{E}$ factors, there remains a rapidly converging sum. The
constant $M$ in general depends on propagation and polarization direction,
and can also be negative. The value~(\ref{eq:M_Madelung}) is for
the propagation along an axis of a cubic crystal. The final expression
for the stress tensor is

\begin{align}
\bar{\sigma}_{33} & =\tfrac{\epsilon_{0}}{4}\left|E_{2}^{\mathrm{loc}}\right|^{2}\left(\left(1-\tfrac{\gamma}{3}\right)^{2}\left(n^{2}+1\right)+\gamma^{2}M\right)\label{eq:sigma_33}\\
 & =\tfrac{\epsilon_{0}}{4}\left|E_{2}^{\mathrm{avg}}\right|^{2}\left(n^{2}+1+M\left(n^{2}-1\right)^{2}\right).\nonumber 
\end{align}
The similar calculation of the transverse stress tensor leads to
\begin{equation}
\bar{\sigma}_{11}=-\bar{\sigma}_{22}=-\tfrac{\epsilon_{0}}{4}\left|E_{2}^{\mathrm{avg}}\right|^{2}\left(n^{2}-1-M\left(n^{2}-1\right)^{2}\right).\label{eq:sigma_1122}
\end{equation}
We now consider a plane wave with a wave front. Into an undisturbed
interval of length $\ell$ (in a time $\ell n$) flows the momentum
$\bar{\sigma}_{33}\ell n=\overline{\Pi_{3}^{\mathrm{tot}}}\ell$,
which gives the expression for the average total momentum density
$\overline{\Pi_{3}^{\mathrm{tot}}}$ listed in table~\ref{tab:MomDensity}.
The mechanical momentum density $\overline{\Pi_{3}^{\mathrm{mech}}}$,
also listed in the table, follows by subtracting the electromagnetic
momentum density~(\ref{eq:PI_em}). The results are most easily
compared with other values as momentum per energy~(\ref{eq:u_tot}),
for example $\overline{\Pi_{3}^{\mathrm{tot}}}/\overline{u^{\mathrm{tot}}}.$
The ratios differ from the usually assumed values. This is discussed
further below.

\begin{table}
\centering{}%
\begin{tabular}{ccc}
\hline 
Quantity & This work & Conventional\tabularnewline
\hline 
$\overline{u^{\mathrm{tot}}}$ & $\tfrac{\epsilon_{0}}{2}\left|E_{2}^{\mathrm{avg}}\right|^{2}n^{2}$ & $\tfrac{\epsilon_{0}}{2}\left|E_{2}^{\mathrm{avg}}\right|^{2}n^{2}$\tabularnewline
$\overline{\Pi_{3}^{\mathrm{em}}}/\overline{u^{\mathrm{tot}}}$ & $1/n$ & $1/n$\tabularnewline
$\overline{\Pi_{3}^{\mathrm{tot}}}/\overline{u^{\mathrm{tot}}}$ & $\tfrac{1}{2}\left(n+\tfrac{1}{n}+M\left(n^{2}-1\right)^{2}/n\right)$ & $n$\tabularnewline
$\overline{\Pi_{3}^{\mathrm{mech}}}/\overline{u^{\mathrm{tot}}}$ & $\tfrac{1}{2}\left(n-\tfrac{1}{n}+M\left(n^{2}-1\right)^{2}/n\right)$ & $n-1/n$\tabularnewline
\hline 
\end{tabular}\caption{\label{tab:MomDensity}Average energy density $\overline{u^{\mathrm{tot}}}$
and electromagnetic, total and mechanical momentum per energy. }
\end{table}

The momentum density $\mu v_{3}\sim\overline{u^{\mathrm{tot}}}/c$
of matter with a mass density $\mu\gg\overline{u^{\mathrm{tot}}}/c^{2}$
and a velocity $v_{3}$ can be significant, while its \emph{momentum
flow} $\mu v_{3}^{2}$ is negligible compared to the electromagnetic
momentum flow $\bar{\sigma}_{33}$. Consequently, the momentum flow
of a wave train is given by (\ref{eq:sigma_33}), independently of
whether matter is at rest or has been accelerated by the wave front.
The kinetic energy also is negligible.

\subsection{Mechanical momentum from Lorentz force}

To get a more intuitive explanation for the mechanical momentum $\overline{\Pi_{3}^{\mathrm{mech}}}$
we now calculate the Lorentz force $j\times B$ acting on the atoms
in the front of a light signal. To allow for the construction of propagating
wave packets we now keep the time dependence $e^{-i\omega t}.$

A key observation, not restricted to the model, is that the magnetic
field~(\ref{eq:B_x_3D}) contains only a $q=0$ component. This reflects
the fact that magnetic fields arise only in the far-field zone of
electric dipole radiation~\cite{Jackson1975}, and therefore exhibit
negligible spatial granularity within the medium in the long wavelength
limit. As a result, the spatial integral of the Lorentz force can
be evaluated using the average current density. From Eq.~(\ref{eq:j_Fourier}),
Eq.~(\ref{eq:E_avg}) and Eq.~(\ref{eq:n2_3D}) it follows

\[
j_{2}^{\mathrm{avg}}=\tfrac{\partial}{\partial t}\epsilon_{0}\left(n^{2}-1\right)E_{2}^{\mathrm{avg}}e^{ik_{3}\left(x_{3}-t/n\right)}.
\]
Wave packets are superpositions of plane waves with wavevectors centered
around some wavevector $k$. A convenient parametrization has a slowly
varying function $E_{2}^{\mathrm{avg}}\left(x_{3}-t/n\right)$ instead
of a constant amplitude $E_{2}^{\mathrm{avg}}$. If $E_{2,p}^{\mathrm{avg}}$
denotes the Fourier transform of $E_{2}^{\mathrm{avg}}\left(x_{3}\right)$
then 

\begin{align}
g\left(x_{3}-t/n\right) & \equiv\Re E_{2}^{\mathrm{avg}}\left(x_{3}-t/n\right)e^{ik_{3}\left(x_{3}-t/n\right)}\label{eq:WavePacket_E}\\
 & =\Re\int\tfrac{\mathrm{d}p}{2\pi}E_{2,p-k_{3}}^{\mathrm{avg}}e^{ip\left(x_{3}-t/n\right)}.\nonumber 
\end{align}
The magnetic field~(\ref{eq:B_x_3D}) becomes

\[
B_{1}=-ng\left(x_{3}-t/n\right).
\]
The average Lorentz force density thus is

\[
\overline{f_{3}}\left(x_{3},t\right)=\tfrac{\epsilon_{0}}{2}\left(n^{3}-n\right)\tfrac{\partial}{\partial t}\overline{g^{2}\left(x_{3}-t/n\right)}.
\]
Integrating over time and averaging locally over space (or time) yields
the corresponding mechanical momentum density 
\begin{align}
\overline{\Pi_{3}^{\mathrm{lorentz}}} & =\tfrac{\epsilon_{0}}{2}\left(n^{3}-n\right)\left.\overline{g^{2}\left(x_{3}-t/n\right)}\right|_{-\infty}^{t}\label{eq:PI_mech_Lorentz}\\
 & =\tfrac{\epsilon_{0}}{4}\left(n^{3}-n\right)\left|E_{2}^{\mathrm{avg}}\right|^{2},\nonumber 
\end{align}
where we have assumed $g\left(\infty\right)=0$ for a signal with
a wave front. Eq.~(\ref{eq:PI_mech_Lorentz}) explains the first
part of $\overline{\Pi_{3}^{\mathrm{mech}}}$ in table~\ref{tab:MomDensity}.
The result~(\ref{eq:PI_mech_Lorentz}) is generic, it does not depend
on the details of the model.

\subsection{Mechanical momentum from Coulomb force}

Within a wave front the dipole moments of the atoms increase. This
leads to an electrostatic force in propagation direction. Every atom
first is in an environment with no polarization, then in an environment
with a polarization gradient and then the polarization is constant. 

A linear distribution of dipole moments $p\left(x\right)=p_{0}+\left(\Delta p_{0}/\Delta x_{3}\right)x_{3}$
with a constant gradient generates a force in $x_{3}$ direction on
the atom at the origin,

\begin{align}
F_{3}^{\mathrm{coulomb}} & =\tfrac{-p_{0}}{4\pi\epsilon_{0}}\sum_{x\in\left(\mathbb{Z}a\right)^{3}\setminus0}p\left(x\right)\left(\tfrac{3}{\left|x\right|^{5}}-\tfrac{15x_{2}^{2}}{\left|x\right|^{7}}\right)x_{3}\Lambda\left(x_{3}\right)\label{eq:F3_Coul}\\
 & =-\tfrac{1}{\epsilon_{0}}p_{0}\tfrac{\Delta p_{0}}{\Delta x_{3}}a^{-3}M',\nonumber \\
M' & =\tfrac{1}{4\pi}\sum_{x\in\mathbb{Z}^{3}\setminus0}\left(\tfrac{3}{\left|x\right|^{5}}-\tfrac{15x_{2}^{2}}{\left|x\right|^{7}}\right)x_{3}^{2}\Lambda\left(x_{3}\right)\label{eq:M2}\\
 & \cong0.1716\cong M.\nonumber 
\end{align}
The constant $M'$ is another dimensionless Madelung constant. The
sum is only conditionally convergent, and it is crucial to use the
right anisotropic cutoff, represented by the function $\Lambda\left(x_{3}\right)=\theta\left(\lambda^{2}-x_{3}^{2}\right)$.
The oscillating amplitude of the dipoles in propagation direction
restricts the sum to a finite slice $-\lambda<x_{3}<\lambda$ (the
wave front extends over many wavelengths). The exact form of the cutoff
function is irrelevant, the sum rapidly converges in $x_{3}$ direction.
There is no such restriction in the transverse direction. The value
given in Eq.~(\ref{eq:M2}) is the result of an evaluation with $\lambda=4$
or $\lambda=5$ on a computer, the (approximate) relation to the Madelung
constant $M$ from Eq.~(\ref{eq:M_Madelung}) is as expected. We
have not attempted to independently prove $M'=M$. 

The sum over the force~(\ref{eq:F3_Coul}) along the signal front
gives the momentum transferred to the atoms,
\begin{align}
\Pi_{3}^{\mathrm{coulomb}} & =-\tfrac{1}{\epsilon_{0}}\sum_{p_{0}}p_{0}\Delta p_{0}a^{-4}M\label{eq:PI_mech_Coulomb}\\
 & \cong\tfrac{1}{2\epsilon_{0}}p_{0}^{2}a^{-4}M=\tfrac{\epsilon_{0}}{2}\gamma^{2}\left|E^{\mathrm{loc}}\right|^{2}M\nonumber \\
 & =\tfrac{\epsilon_{0}}{4}\left|E^{\mathrm{avg}}\right|^{2}M\left(n^{2}-1\right)^{2}n.\nonumber
\end{align}
Here we have used Eq.~(\ref{eq:Dip_Moment}). This explains the second
part of $\overline{\Pi_{3}^{\mathrm{mech}}}$ in table~\ref{tab:MomDensity}.
The Coulomb contribution to the mechanical momentum in general depends
on the direction of propagation and polarization.

There is a simpler model with polarizable thin layers instead of atoms~\cite{Dengler2025},
which can be exactly solved with a Bloch Ansatz. The results agree
with table~\ref{tab:MomDensity}, except that there is no contribution
from the Coulomb force. This is plausible since no charge accumulates
in the layers when the layers are polarized. And since the electric
field parallel to the layers is continuous there also is no Clausius-Mossotti
relation.

\section{Relation to Peierls's work}

We have derived the same total momentum per energy
\begin{equation}
\overline{\Pi_{3}^{\mathrm{tot}}}/\overline{u^{\mathrm{tot}}}=\tfrac{1}{2}\left(n+1/n+M\left(n^{2}-1\right)^{2}/n\right)\label{eq:PI_tot_u_tot}
\end{equation}
with $M\cong0.1716$ for a propagation in the direction of an axis
of the cubic crystal with two different methods. The conventionally
accepted value is $n$, and there evidently is a problem.

There is little doubt in the DDA solution~(\ref{eq:E_x_3D}, \ref{eq:B_x_3D}).
Neglecting the displacement of atoms is well motivated, and the (small)
atom radius $R$ does not enter the result. The electric field constitutes
a complete set of degrees of freedom, the magnetic field follows by
integrating the law of induction over time, the particle momentum
follows by integrating the Lorentz and Coulomb force acting on atoms
over time.

As emphasized above, several features of the solution\ (\ref{eq:E_x_3D},
\ref{eq:B_x_3D}) are quite universal. The magnetic field originates
from the far zone of electric dipole radiation, is non-granular, and
combines with the universal (non-granular, average) component of the
electric field in both the Poynting vector and the Lorentz force.
Microscopic details enter only through the Madelung constant $M$.

It is therefore not surprising to find essentially the same result~(\ref{eq:PI_tot_u_tot})
in the work of R. Peierls~\cite{Peierls1976,Peierls1977}, who also
uses the fact that the magnetic field is non-granular. His Eq.~(2.12)
for total momentum agrees with our Eq.~(\ref{eq:PI_tot_u_tot}) except
for his constant $-\sigma=-0.2$ for glasses and liquids instead of
our $M\cong0.1716$ for the propagation along a crystal axis of a
cubic crystal. A derivation of Peierls's constant can be found in
appendix~\ref{subsec:M2_liq}. 

It remains to resolve the discrepancy with experiments, which tend
to confirm the conventional values. Instead of providing a summary
of experimental work here we refer to the reviews~\cite{Brevik1979,Pfeifer_2007,BarnLoud2010}.
The signal momentum usually is small and results sometimes are controversial.

A unique result for liquids is due to Jones and Leslie~\cite{Jones_Leslie_1978}.
The authors describe an elaborate repetition of an experiment by Jones
and Richards~\cite{Jones_Rich_1954}, in which a torsion balance
with a mirror is submerged in air or a liquid. A laser ray is asymmetrically
directed to the mirror, and the angular deviation is measured. For
various liquids the authors find a momentum proportional to $n$ with
an accuracy of $\unit[0.05]{\%}$. For the used signal with a power
$15\unit{mw}$ in a cross section of $\left(\unit[0.6]{mm}\right)^{2}$
the displacement~(\ref{eq:Shift_Estimate}) of the atoms clearly
is negligible.

\section{Reflection at a mirror}

Since all equations are linear incoming and reflected waves~(\ref{eq:E_x_3D},
\ref{eq:B_x_3D}) can be superimposed to get the fields of the standing
wave for normal incidence on a mirror at $x_{3}=0$,

\begin{align}
E_{2}^{\mathrm{s}}\left(x\right) & =2iE_{2}^{\mathrm{loc}}\sin\left(k_{3}x_{3}\right)e^{-i\omega t}\left[\left(1-\tfrac{\gamma}{3}\right)\right.\label{eq:E_Stand}\\
 & \qquad\qquad\left.-\gamma\sum_{q\neq0}e^{iqx}\tfrac{qq_{2}}{q^{2}}F\left(qR\right)\right],\nonumber \\
B_{1}^{\mathrm{s}}\left(x\right) & =-2E_{2}^{\mathrm{loc}}n\left(1-\tfrac{\gamma}{3}\right)\cos\left(k_{3}x_{3}\right)e^{-i\omega t}.\label{eq:B_Stand}
\end{align}
The boundary condition is $E_{\Vert}^{\mathrm{s}}=0$ at the mirror.
The electromagnetic stress tensor~(\ref{eq:sigma_33_def}) at $x_{3}\in\left(\mathbb{Z}+\tfrac{1}{2}\right)a$
now becomes  
\begin{align}
\sigma_{33}^{\mathrm{s}} & =2\epsilon_{0}\left(E^{\mathrm{avg}}\right)^{2}\left[n^{2}\cos^{2}\left(k_{3}x_{3}\right)\cos^{2}\left(\omega t\right)\right.\label{eq:sigma_s}\\
 & \qquad+\left.\left(M\left(n^{2}-1\right)^{2}+1\right)\sin^{2}\left(k_{3}x_{3}\right)\sin^{2}\left(\omega t\right)\right].\nonumber 
\end{align}
At the distance $x_{3}=-a/2$ in front of the mirror the electric
field is negligible, and on average there remains

\begin{equation}
\overline{\sigma_{33}^{\mathrm{s,tot}}}=\epsilon_{0}\left(E^{\mathrm{avg}}\right)^{2}n^{2},\label{eq:sigma_s_tot}
\end{equation}
leading to the usual momentum transfer to the mirror. 

The time average of the force acting on the layers is $-a\partial_{3}\overline{\sigma_{33}^{\mathrm{s}}}\propto\sin\left(2k_{3}x_{3}\right)$.
This force tends to deform the medium, but on average deposits no
mechanical momentum in the medium. The effective stress tensor in
the standing wave thus is $\overline{\sigma_{33}^{\mathrm{s,tot}}}$,
independently of whether of the force on the layers (after some time)
is balanced by a mechanical stress or not. It agrees with the electromagnetic
stress tensor~(\ref{eq:sigma_s}) only where the electric field vanishes,
that is for $\sin\left(k_{3}x_{3}\right)=0$. 

Now consider a signal of length $\ell$ entering from $x_{3}<0$.
The stress tensor in the propagating signal is constant in space and
identical to the electromagnetic stress tensor~(\ref{eq:sigma_33})
calculated in the vacuum between the atoms. A standing wave then exists
for a time $n\ell$. In this time it transfers momentum $\overline{\sigma_{33}^{\mathrm{s,tot}}}n\ell$
to the mirror, and the opposite momentum to the propagating signal.
The mismatch of the stress tensors between propagating and standing
waves means that mechanical momentum is deposited at the (moving)
boundary of the standing wave. The fact that the standing wave expands
from $x_{3}=0$ to $-\ell/2$ and then retracts to $x_{3}=0$ complicates
the picture. A simple solution is to consider the momentum in the
interval $-\ell/2\leq x_{3}<0$. The incoming signal adds its momentum
$\bar{\sigma}_{33}n\ell$ at $-\ell/2$, the outgoing signal removes
its momentum $-\bar{\sigma}_{33}n\ell$ also at $-\ell/2,$ and the
standing wave adds its momentum near $x_{3}=0.$ 

Such considerations of course are not new. An equivalent explanation
was presented by Peierls based on calculations for liquids at the
level of continuum electrodynamics~\cite{Peierls1976}. He concluded
that two waves, an incoming and a reflected one, are always required,
along with a refractive index different from one, for such additional
momentum deposition to occur. No such interference arises when a signal
gets absorbed or when a signal enters a dielectric medium from vacuum.
The rather complicated extension~\cite{Peierls1977} of Peierls's
calculation to oblique incidence with polarization parallel to the
plane of incidence leads to results not compatible with experiment~\cite{Jones_Leslie_1978}.
We claim that these calculations are wrong. 

\begin{figure}
\centering{}\includegraphics{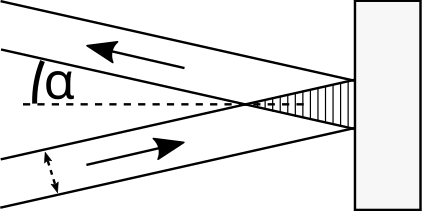}\caption{\label{fig:ObliqueIncidence}A long wave train of finite width reflected
at a mirror. The standing wave in the hatched triangle transfers momentum
to the mirror and to the boundaries with the propagating waves. The
mismatch of the stress tensors leads to a deposition of mechanical
momentum at the boundaries.}
\end{figure}

The above considerations based on constant stress tensors in standing
and propagating waves can directly be extended to the case of oblique
incidence, see figure~(\ref{fig:ObliqueIncidence}). The region of
the standing wave now even is stationary, except for a negligible
duration when the fronts of the wave train touch the mirror. As for
perpendicular incidence at the mirror $E_{\Vert}=0$.

For a polarization perpendicular to the plane of incidence, where
also $E_{\bot}=0$, no Lorentz or Coulomb (=dipole) forces act on
the atoms near the mirror. In this case, the magnetic field $B^{\mathrm{avg}}$
alone determines $\sigma_{33},$ leading to the conventional momentum
transfer. 

For oblique incidence with electric field in the plane of incidence,
the field has a component perpendicular to the mirror. One could begin
complicated calculations using a superposition of plane waves~(\ref{eq:E_x_3D})
and then notice that the boundary condition $E_{\Vert}=0$ is generally
not exactly satisfied because of the dipoles near the mirror's surface.
Such calculations are superfluous. The situation near the mirror is
analogous to that near a conductor in a planar capacitor with a quasi-static
electric field: the field is perpendicular to the conductor, and the
average momentum flow from the electric field is directly determined
by the average field itself.

In all cases the standing wave, a superposition of incoming and outgoing
waves, is invariant under space inversion $x_{3}\rightarrow-x_{3}$,
and on average therefore deposits no mechanical momentum in its interior.
The momentum transferred from the mirror emerges at the border to
the propagating waves. 

\section{Discussion}

For reflection of plane waves at a mirror, the model reproduces the
conventional, experimentally confirmed momentum transfer for all angles
of incidence and all directions of polarization. The model does not
satisfactorily explain the observed momentum transfer ``$2n\cos\alpha"$
to a mirror in the case of finite width signals hitting the mirror
obliquely. In this case the momentum difference is deposited near
the mirror and most of it will reach the mirror by hydrodynamic relaxation
(the time scale in water is $1\mathrm{s}$ for a signal of width $1\mathrm{mm}$).
Apparently something still is missing. Experiments could guide the
theory.

The intrinsic signal momentum~(\ref{eq:PI_tot_u_tot}) could be measured
in an adaptation of the Leslie and Jones experiment, replacing the
mirror with an absorber. In this case there is no interference between
incoming and reflected signals, and the exact solution directly predicts
the momentum transfer. An experimental difficulty would be controlling
the effects of the heat generation, see~\cite{Jones_Leslie_1978}. 

An apparent paradox for continuous or long signals illustrates an
essential aspect of the problem. In long wave trains the moving matter
is slowed down hydrodynamically, apparently indicating that the residual
electromagnetic momentum~(\ref{eq:PI_em}) (``$1/n$'') is passed
to the absorber. This conclusion, however, is incorrect, as the stress
tensor is still given by (\ref{eq:sigma_33}), see section~\ref{subsec:sigma33}.
The ``missing'' momentum comes from the medium: the trailing edge
of the signal now leaves backward-moving matter. The opposite momentum
is transferred to the signal. For short signals, in contrast, matter
as in~figure(\ref{fig:Signal}) is at rest when the signal has passed,
and the transferred momentum equals the signal's momentum.

It also would be of interest to measure the transverse stress tensor
(\ref{eq:sigma_1122}). Although less significant dynamically, it
shares the same origin and form as the longitudinal component and
likewise couples to matter.

In a liquid, the transverse stress tensor induces a $2$d quadrupole
stokeslet around a signal of finite cross section. The flow is polarization
dependent, directed outward along $\pm E$ and inward along $\pm B$.
In water, the velocity at $1\,\mathrm{mm}$ from a $10\,\mathrm{mW}$
source is of order $10^{-3}\mathrm{mm}/\mathrm{s}$, decaying like
$1/d$ with distance $d$. The velocity is larger in liquids with
a lower dynamic shear viscosity. 

\section{Conclusion}

In situations where quantum mechanics does not play a special role
it must be possible to describe the propagation of macroscopic electromagnetic
signals with long wavelengths in dielectric media at the level of
classical electrodynamics and classical mechanics. 

As a step in this direction we have presented a simple model allowing
an exact solution in the limit of long wavelengths. Treating the atoms
as point dipoles is only a minor simplification, as most results do
not depend on the size of the dipoles. The resulting momentum of
light in a dielectric medium is the average of the Minkowski and Abraham
momenta, with an additional contribution from Coulomb forces. The
contribution from dipole forces to the mechanical momentum also appears
in Peierls\textquoteright s theory from nearly 50 years ago.

A multitude of experiments seem to confirm an intrinsic momentum \textquotedblleft $n$\textquotedbl .
Their detailed analysis is beyond the scope of this work. The customary
interpretation of the signal momentum as a context-dependent mixture
of electrodynamic and mechanical momentum is unscientific. We thus
emphasize that a consistent microscopic model is essential. The value
of such models is not negated by contradictory experiments; rather,
they can guide future theoretical work.

\appendix

\section{Appendix}

\subsection{Conventions}

The conventions for the Fourier transform of periodic functions on
a cubic lattice with spacing $a$ are
\[
f\left(x\right)=a^{-3}\sum\limits_{q\in\left(2\pi\mathbb{Z}/a\right)^{3}}e^{iqx}f_{q},\quad f_{q}=\int_{a^{3}}\mathrm{d}^{3}xe^{-iqx}f\left(x\right).
\]
When using complex values for physical quantities what is meant is
the real part. Averages of sinusoidal bilinear quantities over time
and a cell can be calculated with the formula
\begin{align*}
a^{-3}\int_{a^{3}}\mathrm{d}^{d}x\overline{\Re f\left(x\right)\Re g\left(x\right)} & =\tfrac{1}{2}a^{-3}\int\mathrm{d}^{d}x\Re\left(fg^{*}\right).
\end{align*}

\subsection{Shape function $F$}

The Fourier transforms of charge and current density contain the function
\begin{align}
F\left(qR\right) & =\tfrac{1}{V_{R}}\int_{\left[-a/2,a/2\right]^{3}}\mathrm{d}^{3}xe^{-iqx}\theta\left(R-\left|x\right|\right)\label{eq:F_s_Def}\\
 & =3\left(\sin\left(qR\right)-qR\cos\left(qR\right)\right)/\left(qR\right)^{3},\nonumber
\end{align}
where $\theta$ denotes the step function. The integral domain has
be chosen symmetrically around the origin.

\subsection{Electric field energy}

\label{subsec:u_em_calc}Inserting (\ref{eq:E_x_3D}) into the expression
(\ref{eq:u^elec}) for the average electric energy density leads to
\begin{align*}
\overline{u^{\mathrm{elec}}} & =\tfrac{\epsilon_{0}}{4}\left|E_{2}^{\mathrm{loc}}\right|^{2}\left[\left(1-\tfrac{\gamma}{3}\right)^{2}\right.\\
 & \qquad\qquad\left.+\gamma^{2}\sum\limits_{q\in\left(2\pi\mathbb{Z}/a\right)^{3}\setminus0}\tfrac{q_{2}^{2}}{q^{2}}F^{2}\left(qR\right)^{2}\right]\\
 & =\tfrac{\epsilon_{0}}{4}\left|E_{2}^{\mathrm{loc}}\right|^{2}\left(\left(1-\tfrac{\gamma}{3}\right)^{2}-\tfrac{\gamma^{2}}{3}+\tfrac{\gamma^{2}}{3}\tfrac{a^{2}}{V_{R}}\right).
\end{align*}
Here we have used

\begin{align*}
\sum\limits_{q\in\left(2\pi\mathbb{Z}/a\right)^{3}\setminus0}\tfrac{q_{2}^{2}}{q^{2}}F^{2}\left(qR\right) & =\tfrac{1}{3}\sum\limits_{q\in\left(2\pi\mathbb{Z}/a\right)^{3}}F^{2}\left(qR\right)-\tfrac{1}{3}\\
 & =\tfrac{a^{3}}{3V_{R}}-\tfrac{1}{3}.
\end{align*}
The remaining sum has been performed by inserting Eq.~(\ref{eq:F_s_Def})
and using $\sum_{q}e^{-iqx}=a^{3}\sum_{m}\delta\left(x-x_{m}\right)$,
which leads to the integral,

\begin{align}
\sum_{q}F^{2}\left(qR\right)= & a^{3}V_{R}^{-2}\int\mathrm{d}^{3}x\theta^{2}\left(R-\left|x\right|\right)=a^{3}/V_{R}.\label{eq:sum_int_F2}
\end{align}

\subsection{A formula for a sum over wavevectors}

\label{subsec:Sum_cosh_formula}The electromagnetic stress tensor
(\ref{eq:sigma_33}) in a plane between the atoms was evaluated with
the formula

\begin{equation}
f_{Q}\left(z\right)=\sum_{q\in2\pi\mathbb{Z}}\tfrac{e^{iqz}}{Q^{2}+q^{2}}=\tfrac{1}{4}\tfrac{\cosh\left(Q\left(z-\tfrac{1}{2}\right)\right)}{\tfrac{Q}{2}\sinh\left(\tfrac{Q}{2}\right)},\label{eq:sum_cosh}
\end{equation}
valid for $Q\neq0$ in the interval $0\leq z\leq1$. Symmetry properties
are $f_{Q}\left(z\right)=f_{Q}\left(-z\right)$ and $f_{Q}\left(z\right)=f_{Q}\left(z+1\right)$.
To verify the formula it suffices to check that $f_{Q}\left(z\right)$
satisfies the differential equation 

\[
\left(\partial_{z}^{2}-Q^{2}\right)f_{Q}\left(z\right)=-\sum_{m\in\mathbb{Z}}\delta\left(z-m\right)
\]
and the symmetry conditions. The function $f_{Q}\left(z\right)$ is
a linear combination of the solutions $e^{\pm Qz}$ of the homogeneous
differential equation. The inhomogeneous equation is satisfied because
of $f_{Q}'\left(1\right)-f_{Q}'\left(0\right)=1$.

\subsection{Madelung constant for liquids}

\label{subsec:M2_liq}Peierls's value $M^{\mathrm{liq}}=-1/5$ for
the Madelung constant of liquids and glasses follows from Eq.~(\ref{eq:F3_Coul})
and Eq.~(\ref{eq:M2}) by replacing the sum over lattice points with
an integral $N\int_{\left|x\right|>D}\mathrm{d}^{3}x\ldots$, where
$N$ is the average atom density and $D$ the atom diameter,
\begin{equation}
M^{\mathrm{liq}}=\tfrac{1}{4\pi}\int_{\left|x\right|>D}\mathrm{d}^{3}x\left(\tfrac{3}{\left|x\right|^{5}}-\tfrac{15x_{2}^{2}}{\left|x\right|^{7}}\right)x_{3}^{2}=-\tfrac{1}{5}.\label{eq:M_liq}
\end{equation}
The volume $a^{3}$ in the dipole moment of an atom (\ref{eq:Dip_Moment})
is to be replaced with $1/N$. The constant $M^{\mathrm{liq}}$ measures
the force exerted by a continuum of dipoles at $\left|x\right|>D$
with a dipole moment gradient in $x_{3}$ direction on the dipole
at the origin. All dipoles are oriented in $x_{2}$ direction. There
is no contribution from $\left|x_{3}\right|>D$, uniformly polarized
layers do not exert a force on the dipole at the origin. All integrals
in (\ref{eq:M_liq}) are algebraic, the diameter $D$ drops out. Numerically
we find that $M^{\mathrm{liq}}=-1/5$ agrees with the expression (\ref{eq:M2})
for a cubic crystal averaged over all orientations. 

\bigskip{}

\bibliographystyle{habbrv}
\bibliography{Other}

\end{document}